\documentclass{article}

\usepackage{PRIMEarxiv}
\usepackage{authblk}
\usepackage[utf8]{inputenc} 
\usepackage[T1]{fontenc}    
\usepackage{hyperref}       
\usepackage{url}            
\usepackage{booktabs}       
\usepackage{amsfonts}       
\usepackage{nicefrac}       
\usepackage{microtype}      
\usepackage{lipsum}
\usepackage{fancyhdr}       
\usepackage{graphicx}       
\graphicspath{{media/}}     

\usepackage{cite}

\usepackage{hyperref}
\hypersetup{
  colorlinks=true,
  linkcolor=black,
  citecolor=black,
  urlcolor=black
}
\title{Using BOLD-fMRI to compute the Respiration Volume per Time (RTV) and Respiration Variation (RV) with Convolutional Neural Networks (CNN) in the Human Connectome Development Cohort
\thanks{2023 International Society of Magnetic Resonance in Medicine.  Toronto, Canada,  June 2-9.  Abstract Number 2709}
}

\author[1-5]{Abdoljalil Addeh}
\author[1-5]{Fernando Vega}
\author[7]{Rebecca J Williams}
\author[8]{Ali Golestani}
\author[4-6]{\\G. Bruce Pike}
\author[1-5]{M. Ethan MacDonald}

\affil[1]{Department of Biomedical Engineering, Schulich School of Engineering, University of Calgary, Calgary, AB, Canada}
\affil[2]{Department of Electrical \& Software Engineering, Schulich School of Engineering, \protect\\University of Calgary, Calgary, AB, Canada}
\affil[3]{Alberta Children’s Hospital Research Institute, University of Calgary, Alberta, Canada}
\affil[4]{Hotchkiss Brain Institute, Cumming School of Medicine, University of Calgary, Calgary, AB, Canada}
\affil[5]{Department of Radiology, Cumming School of Medicine, University of Calgary, Calgary, AB, Canada, }
\affil[6]{Department of Clinical Neurosciences, Cumming School of Medicine, University of Calgary, Calgary, AB, Canada }
\affil[7]{Faculty of Health, Charles Darwin University, Australia}
\affil[8]{Department of Medical Physics, Alberta Heath Services, Calgary, AB, Canada}

\begin{document}
\maketitle
\section*{Synopsis}
In many fMRI studies, respiratory signals are unavailable or do not have acceptable quality.  Consequently, the direct removal of low-frequency respiratory variations from BOLD signals is not possible. This study proposes a one-dimensional CNN model for reconstruction of two respiratory measures, RV and RVT.  Results show that a CNN can capture informative features from resting BOLD signals and reconstruct realistic RV and RVT timeseries.  It is expected that application of the proposed method will lower the cost of fMRI studies, reduce complexity, and decrease the burden on participants as they will not be required to wear respiratory bellows.
\section*{Introduction}
Low-frequency variation in breathing rate and depth during functional magnetic resonance imaging (fMRI) scanning can alter cerebral blood flow and consequently, the blood oxygen level-dependent (BOLD) signal. Over the past decade, respiratory response function models \cite{birn2006separating, chang2009relationship, golestani2015mapping, wise2004resting} have shown good performance in modelling the confounds induced by low-frequency respiratory variation. While convolution models perform well, they require the collection of external signals that can be cumbersome and prone to error in the MR environment. 

Respiratory data is not routinely recorded in many fMRI experiments due to the absence of measurement equipment in the scanner suite, insufficient time for set-up, financial concerns, or other reasons \cite{mascarell2020imagen, miller2016multimodal, shafto2014cambridge}. Even in the studies where respiratory timeseries are measured, a large portion of the recorded signals don’t have acceptable quality, particularly in pediatric populations \cite{smith2013resting}. This work proposes a method for estimation of the two main respiratory timeseries used to correct fMRI, including respiration variation (RV) and respiratory volume per time (RVT), directly from BOLD fMRI data. 

\section*{Method}
\begin{figure}
\includegraphics[width=\textwidth]{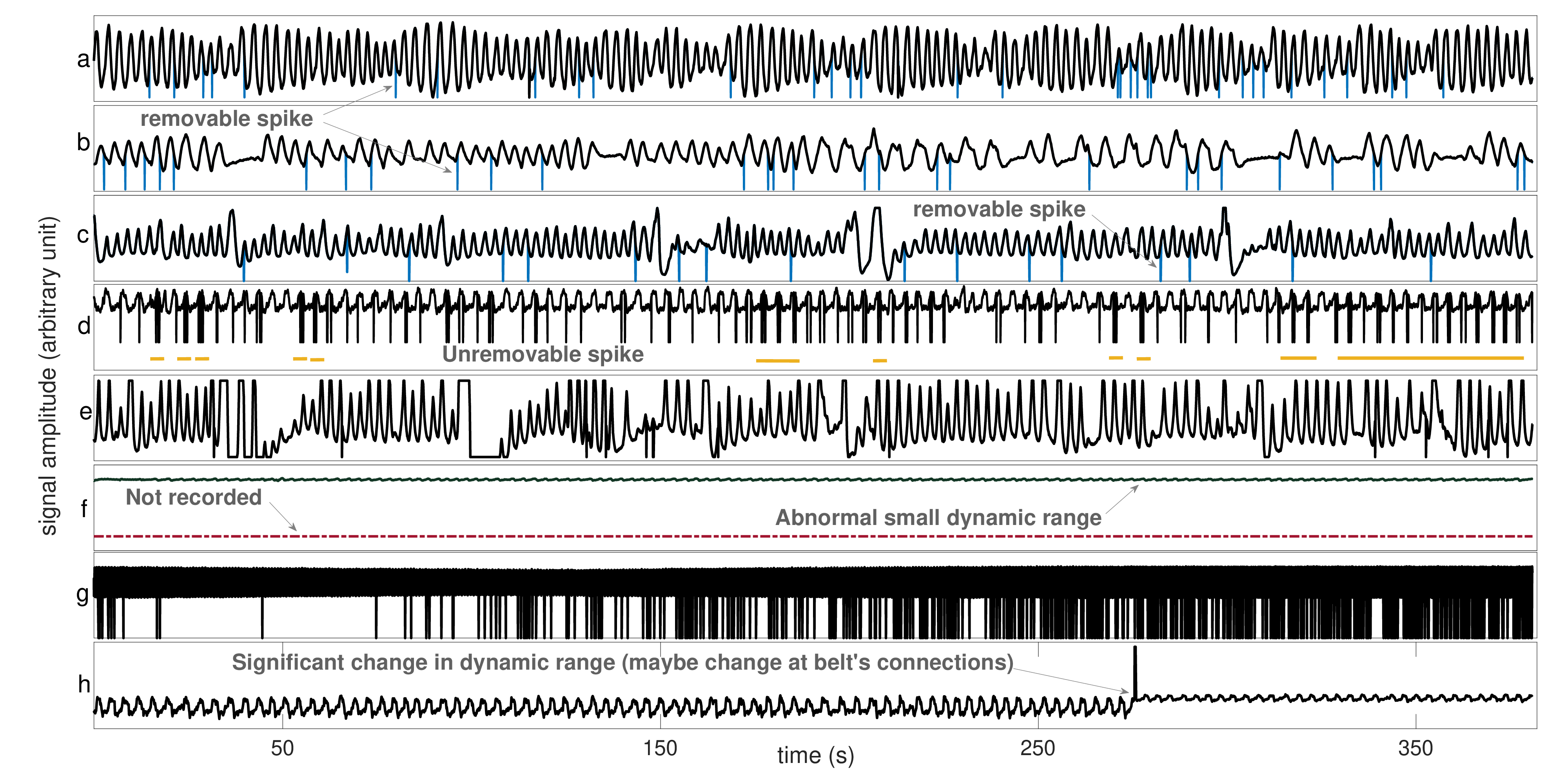}
\caption{Example of usable and unusable respiratory signals. a), b), c) Signal with removable spikes (HCD0271031-2-PA; HCD0008117-2-PA; HCD0001305-2-PA), d) Signal with unremovable spikes marked by color dashes (HCD2335344-2-PA), e) partially recorded signal (HCD1209536-1-AP), f) Not recorded (signal with zero amplitude, HCD0110411-2-AP) and signal varying only in a small range (HCD0694564-1-PA), g) not sampled properly (HCD0146937-1-AP), h) connections changed at a certain point (HCD1240530-1-AP).} 
\label{RespErrors}
\end{figure}

In this study, we used resting-state fMRI scans and the associated respiratory belt traces in Human Connectome Project in Development (HCP-D) dataset (from HCD0001305 to HCD2996590), where participants are children in the age range of 5 and 21 years \cite{somerville2018lifespan, harms2018extending}. From 2451 scans, 306 scans were selected based on the quality of their respiratory signals. Fig. \ref{RespErrors} shows some examples of the usable scans (12.4\%) and unusable scans (87.6\%).

The RVT is the difference between the maximum and minimum belt positions at the peaks of inspiration and expiration, divided by the time between the peaks of inspiration \cite{birn2006separating}, and RV is defined as the standard deviation of the respiratory waveform within a six-second sliding window centered at each time of point \cite{chang2009influence}.  

\begin{figure}
\includegraphics[width=\textwidth]{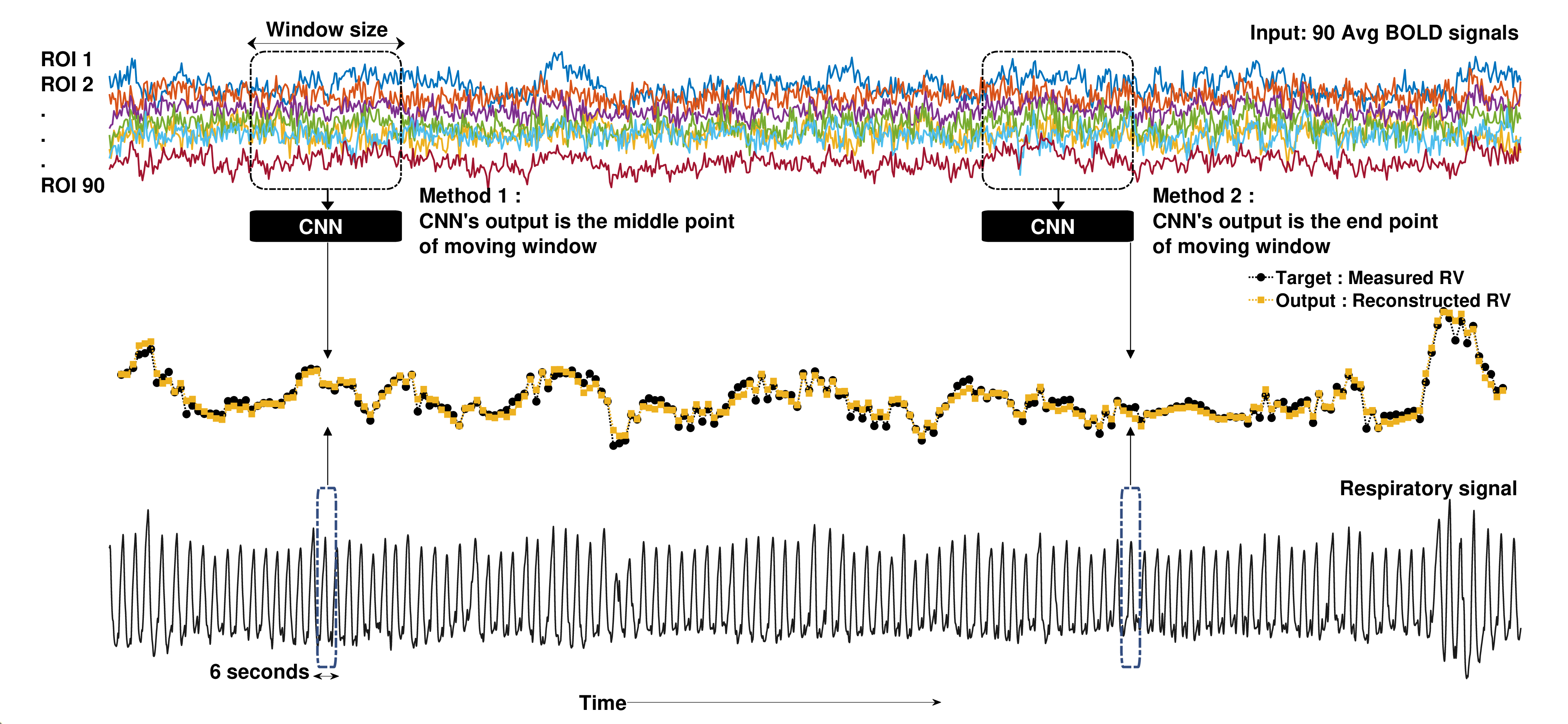}
\caption{Input and output in the proposed method for RV estimation. A respiratory signal is a [152960 $\times$ 1] vector, and RV is a [number of fMRI volumes - window size] vector. Window size determines the length of the input data and number of averaged BOLD signals in each ROI determines the number of channels. Therefore, each input data had a size of [64 $\times$ 90], where 64 is the window size and 90 is the number of ROIs. This broad range for window size is chosen to encompass both immediate and longer-range respiratory changes. Similar system is designed for RVT estimation.} 
\label{Flowchart}
\end{figure}

In the proposed method, a CNN is applied in the temporal dimension of the BOLD time series for RV and RVT reconstruction.  To decrease computational complexity, the average BOLD signal time series from 90 functional regions of interest (ROI) \cite{shirer2012decoding} were used as the main inputs.  Fig. \ref{Flowchart} shows the model input and output for RV estimation.  For each RV point, BOLD signals centered at the RV point, covering 32 TRs before and after were used as the input.  The output of the model can be defined as any point of the moving window. In this paper, we implemented two approaches: middle point of the window (Method 1 in Figure 2), and end point of the window (Method 2 in Figure 2). In Method 1, CNN can use both past and future information, but in Method 2 it can only use the past information hidden in the BOLD signals. The same procedure is applied for RVT estimation.  

A ten-fold cross-validation strategy was employed to evaluate the robustness of the proposed model.  The performance of each fold was evaluated based on mean absolute error (MAE), mean squared error (MSE), coefficient of determination of the prediction ($R^2$), and dynamic time warping (DTW). 

\section*{Results}
\begin{figure}
\includegraphics[width=\textwidth]{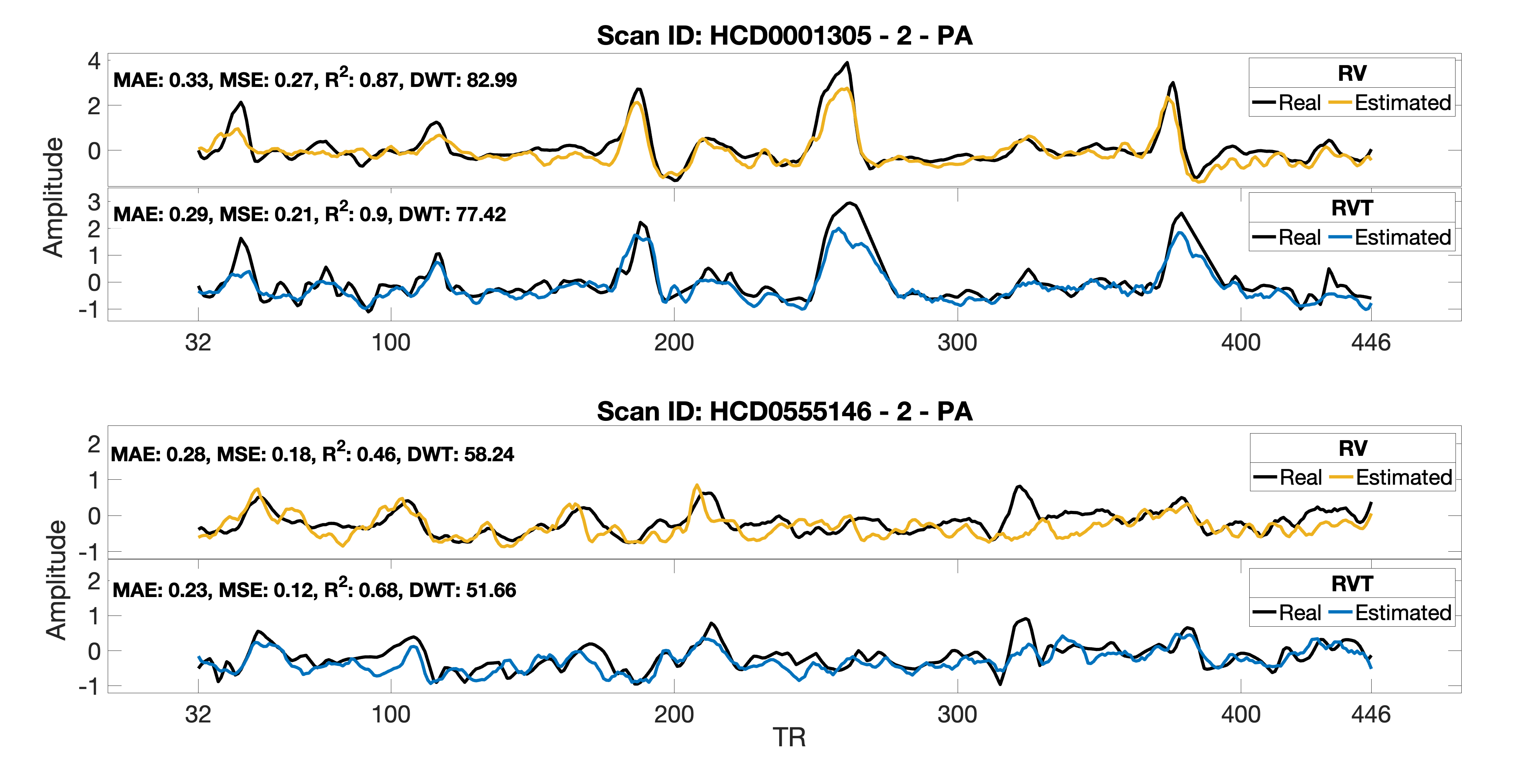}
\caption{Performance of the proposed method when middle point of the moving window with size 64 is selected as CNN’s output. In the HCP-D project, the number of TRs is 478, and using window size 64 leads to loss of 32 points at the beginning and end of the respiratory measures. RV and RVT timeseries with higher values and fluctuations are reconstructed with higher accuracy, and RV and RVT timeseries having nearly constant values are difficult cases for CNN to reconstruct.} 
\label{Perf64}
\end{figure}

Fig. \ref{Perf64} shows the performance of the proposed method considering the middle point of the window as the network’s output on two samples from the test subset.  The trained network can reconstruct the RV and RVT timeseries well, especially when there are big changes in RV and RVT value.  Large changes in RV and RVT can happen when the subject takes a deep breath or if their breathing pattern changes.  Variations in breathing pattern induce a confound to the BOLD signal and the CNN model can use that information to reconstruct the RV and RVT timeseries.  

\begin{figure}
\includegraphics[width=\textwidth]{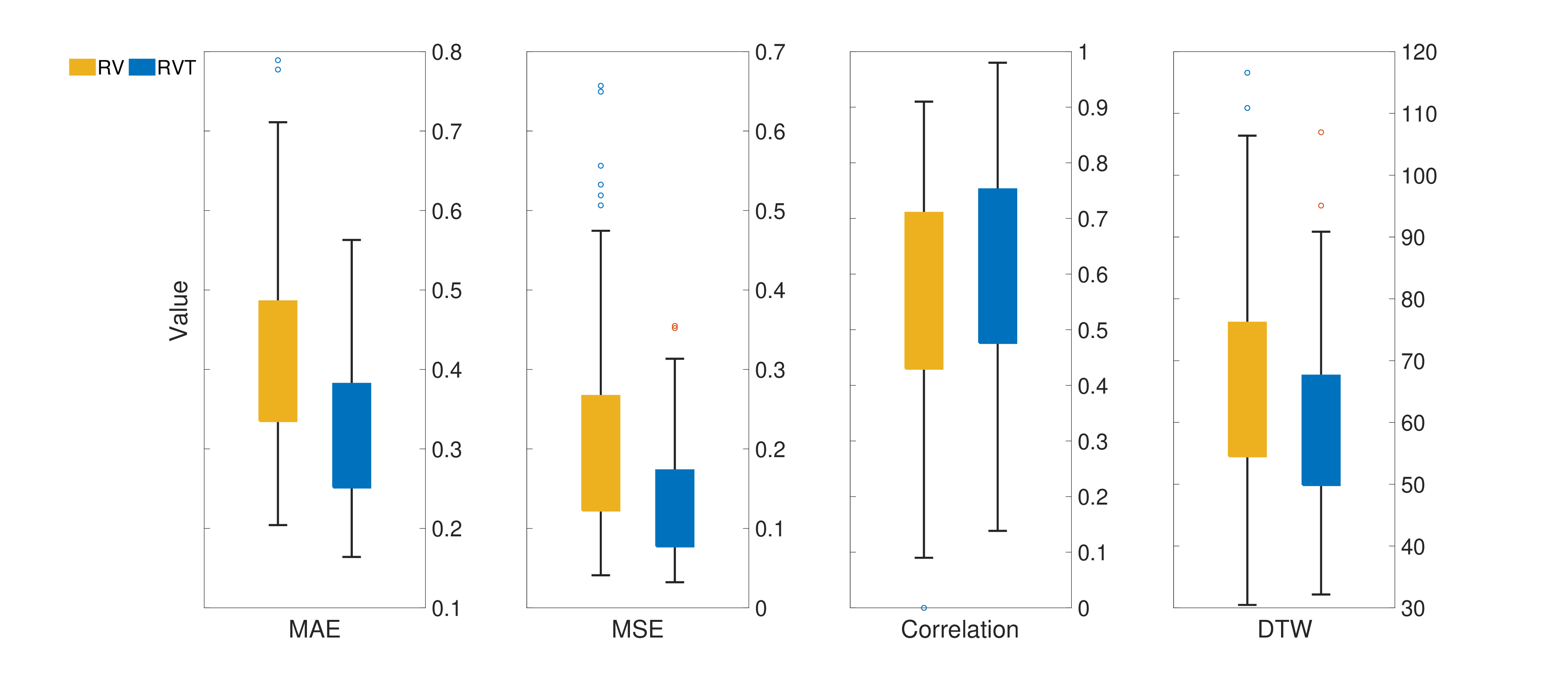}
\caption{Performance of the proposed method shown as box plots of MAE, MSE, correlation, and DTW. Since the MSE squares the errors, it ensures that we don't have any outlier predictions with large errors. As pediatric population is characterized by higher respiration rates and more abnormal breathing patterns, MSE is a more appropriate choice for loss function.  Therefore, we used MSE as the loss function in these experiments. In general, differences among metrics implies that using one loss function for training a machine learning model is not sufficient.} 
\label{boxplots}
\end{figure}

Fig. \ref{boxplots} shows the performance of the CNN considering the middle point of the window as the CNN’s output in terms of MAE, MSE, correlation, and DTW.  As the performance is a function of the variation, no one performance metric is satisfactory.

\begin{figure}
\includegraphics[width=\textwidth]{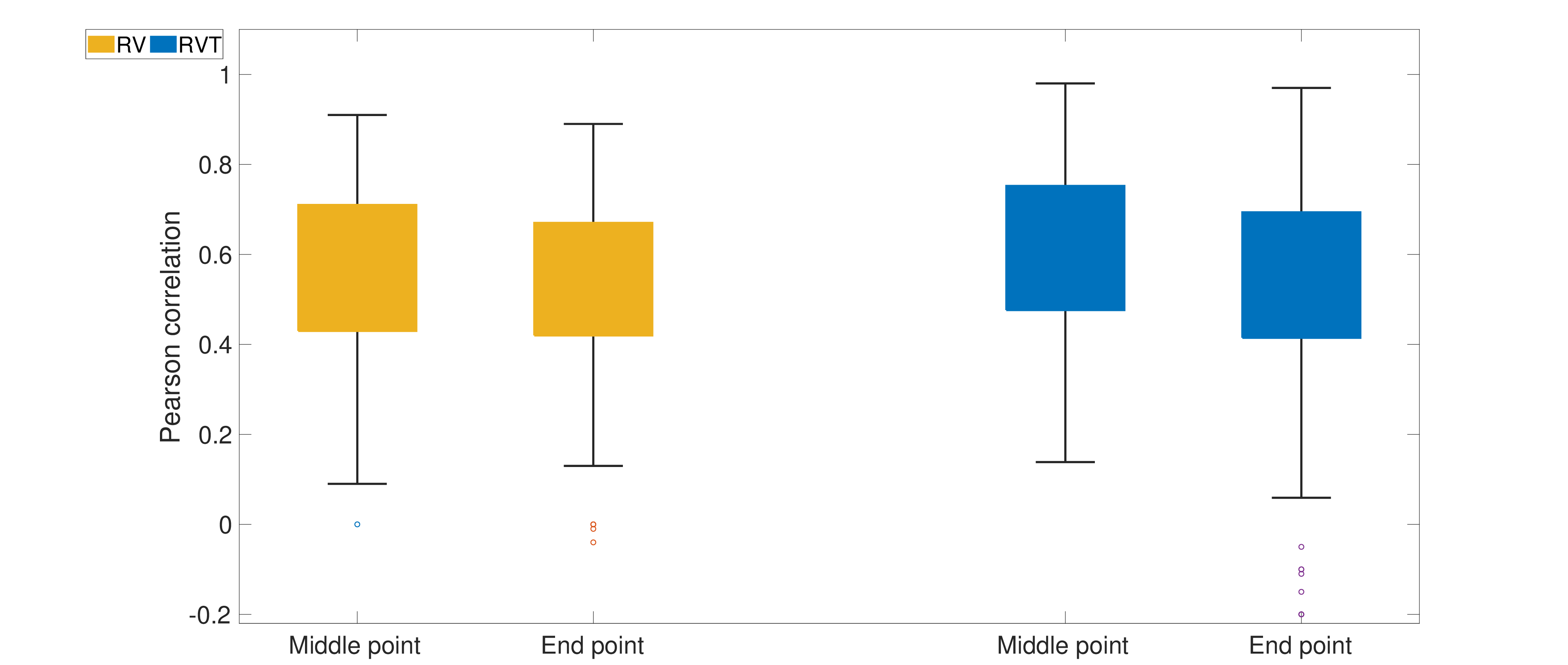}

\caption{Impact of output point location, middle or end point of the moving window, on network’s performance. There is a noticeable difference between two approaches which implies the importance of using input information properly.} 
\label{endpoint}
\end{figure}

Fig. \ref{endpoint} shows the impact of output point location on the network’s performance, whether estimating the point at the middle or end of the window.  The obtained results showed that using both sides, before and after the current breath, leads to better performance. 

\section*{Discussion}
The current work demonstrates the ability to compute both the RV and RVT signals from the fMRI data alone in a pediatric population, which eliminates the need for an external respiratory measurement device and reduces cost. In this paper, we used a window size of 64, but this is an adjustable parameter. Larger window sizes provide more information about baseline breathing rate and depth and enable the model to provide better estimates of variation. An important limitation of larger window sizes is that they result in more time points discarded at the beginning and end of the scan due to edge-effects.  This may limit the minimum duration of scans that can be reasonably reconstructed with the proposed approach, but for longer scans, a larger window might be acceptable.  For fMRI studies with short scan times, the proposed method using a window size of 16 or 32 will be a reasonable choice with a fair performance. There is potential to enrich a large volume of existing resting-state fMRI datasets through retrospective addition of respiratory signal variation information, which is an interesting potential application of the developed method. 

\section*{Acknowledgements:}
The authors would like to thank the University of Calgary, in particular the Schulich School of Engineering and Departments of Biomedical Engineering and Electrical \& Software Engineering; the Cumming School of Medicine and the Departments of Radiology and Clinical Neurosciences; as well as the Hotchkiss Brain Institute, Research Computing Services and the Digital Alliance of Canada for providing resources. The authors would like to thank the Human Connectome Project for making the data available. JA – is funded in part from a graduate scholarship from the Natural Sciences and Engineering Research Council Brain Create.  GBP acknowledges support from the Campus Alberta Innovates Chair program, the Canadian Institutes for Health Research (FDN-143290), and the Natural Sciences and Engineering Research Council (RGPIN-03880).  MEM acknowledges support from Start-up funding at UCalgary and a Natural Sciences and Engineering Research Council Discovery Grant (RGPIN-03552) and Early Career Researcher Supplement (DGECR-00124).

\bibliographystyle{unsrt}
\bibliography{reference.bib}

\begin{thebibliography}{10}

\bibitem{birn2006separating}
Rasmus~M Birn, Jason~B Diamond, Monica~A Smith, and Peter~A Bandettini.
\newblock Separating respiratory-variation-related fluctuations from
  neuronal-activity-related fluctuations in fmri.
\newblock {\em Neuroimage}, 31(4):1536--1548, 2006.

\bibitem{chang2009relationship}
Catie Chang and Gary~H Glover.
\newblock Relationship between respiration, end-tidal co2, and bold signals in
  resting-state fmri.
\newblock {\em Neuroimage}, 47(4):1381--1393, 2009.

\bibitem{golestani2015mapping}
Ali~M Golestani, Catie Chang, Jonathan~B Kwinta, Yasha~B Khatamian, and J~Jean
  Chen.
\newblock Mapping the end-tidal co2 response function in the resting-state bold
  fmri signal: Spatial specificity, test--retest reliability and effect of fmri
  sampling rate.
\newblock {\em Neuroimage}, 104:266--277, 2015.

\bibitem{wise2004resting}
Richard~G Wise, Kojiro Ide, Marc~J Poulin, and Irene Tracey.
\newblock Resting fluctuations in arterial carbon dioxide induce significant
  low frequency variations in bold signal.
\newblock {\em Neuroimage}, 21(4):1652--1664, 2004.

\bibitem{mascarell2020imagen}
Lea Mascarell~Mari{\v{c}}i{\'c}, Henrik Walter, Annika Rosenthal, Stephan
  Ripke, Erin~Burke Quinlan, Tobias Banaschewski, Gareth~J Barker, Arun~LW
  Bokde, Uli Bromberg, Christian B{\"u}chel, et~al.
\newblock The imagen study: a decade of imaging genetics in adolescents.
\newblock {\em Molecular Psychiatry}, 25(11):2648--2671, 2020.

\bibitem{miller2016multimodal}
Karla~L Miller, Fidel Alfaro-Almagro, Neal~K Bangerter, David~L Thomas, Essa
  Yacoub, Junqian Xu, Andreas~J Bartsch, Saad Jbabdi, Stamatios~N Sotiropoulos,
  Jesper~LR Andersson, et~al.
\newblock Multimodal population brain imaging in the uk biobank prospective
  epidemiological study.
\newblock {\em Nature neuroscience}, 19(11):1523--1536, 2016.

\bibitem{shafto2014cambridge}
Meredith~A Shafto, Lorraine~K Tyler, Marie Dixon, Jason~R Taylor, James~B Rowe,
  Rhodri Cusack, Andrew~J Calder, William~D Marslen-Wilson, John Duncan, Tim
  Dalgleish, et~al.
\newblock The cambridge centre for ageing and neuroscience (cam-can) study
  protocol: a cross-sectional, lifespan, multidisciplinary examination of
  healthy cognitive ageing.
\newblock {\em BMC neurology}, 14:1--25, 2014.

\bibitem{smith2013resting}
Stephen~M Smith, Christian~F Beckmann, Jesper Andersson, Edward~J Auerbach,
  Janine Bijsterbosch, Gwena{\"e}lle Douaud, Eugene Duff, David~A Feinberg,
  Ludovica Griffanti, Michael~P Harms, et~al.
\newblock Resting-state fmri in the human connectome project.
\newblock {\em Neuroimage}, 80:144--168, 2013.

\bibitem{somerville2018lifespan}
Leah~H Somerville, Susan~Y Bookheimer, Randy~L Buckner, Gregory~C Burgess,
  Sandra~W Curtiss, Mirella Dapretto, Jennifer~Stine Elam, Michael~S Gaffrey,
  Michael~P Harms, Cynthia Hodge, et~al.
\newblock The lifespan human connectome project in development: A large-scale
  study of brain connectivity development in 5--21 year olds.
\newblock {\em Neuroimage}, 183:456--468, 2018.

\bibitem{harms2018extending}
Michael~P Harms, Leah~H Somerville, Beau~M Ances, Jesper Andersson, Deanna~M
  Barch, Matteo Bastiani, Susan~Y Bookheimer, Timothy~B Brown, Randy~L Buckner,
  Gregory~C Burgess, et~al.
\newblock Extending the human connectome project across ages: Imaging protocols
  for the lifespan development and aging projects.
\newblock {\em Neuroimage}, 183:972--984, 2018.

\bibitem{chang2009influence}
Catie Chang, John~P Cunningham, and Gary~H Glover.
\newblock Influence of heart rate on the bold signal: the cardiac response
  function.
\newblock {\em Neuroimage}, 44(3):857--869, 2009.

\bibitem{shirer2012decoding}
William~R Shirer, Srikanth Ryali, Elena Rykhlevskaia, Vinod Menon, and
  Michael~D Greicius.
\newblock Decoding subject-driven cognitive states with whole-brain
  connectivity patterns.
\newblock {\em Cerebral cortex}, 22(1):158--165, 2012.

\end{thebibliography}
\end{document}